\documentclass[aps,pre,showpacs,amsmath,amssymb,twocolumn,10pt,nofootinbib]{revtex4-1}
\usepackage{amsthm}
\usepackage{dsfont}

\usepackage{slashed}
\usepackage{graphicx}

\newcommand{\lambdab}{{\overline{\lambda}}}
\newcommand{\chib}{{\overline{\chi}}}
\newcommand{\Psib}{{\overline{\Psi}}}
\newcommand{\Lambdab}{{\overline{\Lambda}}}
\newcommand{\psib}{{\overline{\psi}}}
\newcommand{\sigmab}{{\overline{\sigma}}}
\newcommand{\xib}{{\overline{\xi}}}
\newcommand{\Tr}{{\rm Tr\;}}

\def\bec{\begin{center}}
\def\eec{\end{center}}
\def\beq{\begin{equation}}
\def\eeq{\end{equation}}
\def\bea{\begin{eqnarray}}
\def\eea{\end{eqnarray}}

\begin{document}

\title{Dynamical gauge symmetry breaking in strongly coupled lattice theories}

\author{Simon Catterall, Aarti Veernala}
\affiliation{Department of Physics, Syracuse University, Syracuse, NY 13244, USA}

\date\today

\begin{abstract}
We show how a strongly coupled lattice theory consisting of just fermions and 
gauge fields can exhibit a dynamical Higgs mechanism through
the formation of a gauge invariant four fermion condensate.
Furthermore, we argue that this lattice Higgs phase may survive into
the continuum limit.
\end{abstract}

\pacs{05.50.+q, 64.60.A-, 05.70.Fh}

\maketitle
\section{Introduction}

The idea that the Higgs mechanism can occur through the formation of fermionic condensates is an
attractive one when constructing many theories of Beyond Standard Model (BSM) physics and finds application in technicolor, composite Higgs models,
tumbling and grand unification schemes
\cite{TC-intro,TC-intro2, Raby:1979my, gut}. 
Lattice realizations of these scenarios thus potentially give a rigorous setting for understanding 
how non-perturbative dynamics in models without elementary scalars can
spontaneously break gauge symmetries and potentially can give us new tools to analyze such theories.

Of course the terminology
dynamical or spontaneous gauge symmetry breaking is strictly misleading; the gauge symmetry never truly breaks but instead becomes
hidden and the theory enters a Higgs phase 
characterized by a Yukawa-like behavior for the potential between
static sources. Indeed, in a lattice theory, Elitzur's theorem \cite{Elitzur:1975im} guarantees that any
condensate which is not invariant under the gauge symmetry must necessarily have vanishing
expectation value. Instead, the gauge invariant way to understand the operation of the
Higgs mechanism in such theories is that it proceeds via the condensation of a gauge invariant four
fermion operator. The formation of this
four fermion condensate can be thought of as equivalent to the development of
a non-trivial effective potential for $\phi^\dagger \phi$  where $\phi$ is a composite Higgs
field corresponding to  a bilinear fermionic condensate carrying gauge charges.
With this caveat in mind we will often use the
terminology of spontaneous or dynamical gauge symmetry 
breaking when discussing the appearance of
a Higgs phase
in these theories since it is commonly employed in continuum, gauged-fixed discussions
of the Higgs mechanism.

The construction of lattice theories which exhibit such a Higgs phase
has traditionally proven difficult and indeed a folklore has
developed that it is impossible. This belief derives from two well known results; the Vafa Witten
theorem \cite{Vafa:1984xg} which prohibits spontaneous breaking of global vector symmetries 
and the
well known difficulties of constructing non-perturbative lattice chiral gauge theories (see the reviews
\cite{ Poppitz, Luescher, Golterman, Kikukawa} and references therein).
However, there is a loophole in these arguments; one can imagine a lattice theory  describing
Dirac fermions
which develops a condensate which Higgses a {\it vector} gauge
symmetry at non-zero lattice spacing or equivalently strong coupling.
Furthermore there will be no contradiction with the Vafa Witten theorem if this
symmetry reappears as an axial symmetry in the continuum limit.

In this paper we construct 
a lattice gauge theory based on reduced staggered fermions which appears to possess precisely this
structure and we will show via explicit simulations that indeed the resultant lattice theory
develops a Higgs phase  as a result of strong dynamics. Furthermore, since we observe
no phase transitions separating this phase from the weak coupling region we
infer that this lattice
phase likely survives in the continuum limit where we argue the symmetry takes on an axial
character.

We start our construction by considering a free, massless, continuum theory consisting of four Dirac fields. 
We then show how to twist the fermions so that the action can be rewritten
in terms of fermionic matrices. This twisted action can then be decomposed into two independent
pieces corresponding to the projection of the matrix fermions into two
independent components. The key observation is that
the kinetic terms for each of these components may then
be gauged separately under any internal symmetries. This is completely analogous to
the procedure one follows to build chiral gauge theories. Once this is done the
theory no longer admits gauge invariant mass terms which are
{\it also} invariant under the twisted Lorentz symmetry. It is important to recognize, however,
that the theory still contains an equal number of left and right handed fermions in any
given representation of the gauge group and hence it is {\it not} a chiral gauge theory.

The importance of these twisted continuum theories becomes apparent when one uses them
to derive lattice theories which is done in section~\ref{lattice}. In this case each continuum projected
matrix fermion yields a reduced staggered fermion in the lattice theory and the kinetic terms for
these lattice fields are uncoupled as in the continuum. Subsequently gauging these terms under
independent internal symmetries leads to a result, analogous to the continuum, 
that single site fermion bilinear terms  will, in general, break
gauge invariance. 
We use this feature to build interesting models
by assuming that
the gauge interactions factorize into a strong and weakly coupled sector;  the strong interaction
favoring the generation of condensates which can break the weakly coupled gauge symmetries. 
We consider several examples of this symmetry breaking scenario for both global and
local symmetries and show preliminary
numerical results which are consistent with these arguments.


\section{Continuum twisted theory}
\label{twistedtheory}
Consider first a continuum (Euclidean) theory comprising four degenerate {\it massless} Dirac fields $\psi^i$ where the
index runs from $i=1\ldots 4$. The action for
this theory is invariant under the global symmetry $SO_{\rm Lorentz}(4)\times SU_{\rm flavor}(4)$. Let us
focus on the $SO(4)$ subgroup of the flavor symmetry. Under flavor and Lorentz symmetries the fermions transform
like
\beq
\psi_{\alpha\, i}\to L_{\alpha\beta}\psi^{\beta\, j} F^T_{ji}\eeq
Under the action of the diagonal subgroup $SO^\prime(4)={\rm diag}\left(SO_{\rm Lorentz}\times SO_{\rm flavor}(4)\right)$ corresponding to
choosing $F=L$ above
the fermions transform like matrices $\Psi$ and the action may be written as
\beq
S=\int d^4 x\;{\rm Tr}\left(\Psib \gamma_\mu\partial_\mu \Psi\right)\label{start}\eeq
This is the called the twisted representation of the original theory and is only possible when a subgroup of the flavor
symmetry matches the Euclidean Lorentz symmetry\footnote{It is the same representation used in recent
constructions of supersymmetric lattice theories \cite{Catterall:2009it}}.

Now let us define the projectors $P_\pm X=\frac{1}{2}\left(X\pm \gamma_5 X\gamma_5\right)$ for some $4\times 4$ matrix $X$. 
It is easy to show that  $P_+^2=P_+$, $P_-^2=P_-$ and $P_+P_-=P_-P_+=0$ as usual.
Starting from the field $\Psi$ we can define the projected field  $\Psi_-=P_-\Psi$ with the property that  $P_+\Psi_-=0$.
We will assume that  $\Psi$ transforms in some representation of an internal global symmetry group.
In the same way we can introduce a new twisted matrix field $\Lambda$ transforming in some other representation and project it
down to $\Lambda_+=P_+\Lambda$. Using this we can construct a more general action than that given in eqn.~\ref{start} which is
the sum of two independent projected fields carrying different internal quantum numbers.
\beq
S=\int {\rm Tr} (\Psib_+ \gamma_\mu \partial_\mu \Psi_-) +{\rm Tr}(\Lambdab_- \gamma_\mu \partial_\mu \Lambda_+)\label{kinetic}
\eeq
In a chiral basis  these projected matrices take the explicit $2\times 2$ block form
\begin{eqnarray}
\Psib_+ &=&\left(\begin{array}{cc} \psib_L & 0\\0 & \psib_R\end{array}\right)\quad \Lambdab_-=\left(\begin{array}{cc}0 & \lambdab_R\\ \lambdab_L & 0\end{array}\right)\nonumber\\
\Lambda_+ &=&\left(\begin{array}{cc} \lambda_R & 0 \\ 0 & \lambda_L\end{array}\right)\quad\Psi_-=\left(\begin{array}{cc}0 & \psi_R\\ \psi_L & 0\end{array}\right)
\label{decomp}
\end{eqnarray}
The action written in terms of these $2\times 2$ blocks is then
\begin{eqnarray}
S&=&\int d^4 x\; {\rm tr}\left(\psib_L\sigma_\mu\partial_\mu \psi_L+\psi_R\sigmab_\mu\partial_\mu\psi_R \right) \nonumber \\
&+& \int d^4 x\; {\rm tr} \left(\lambdab_L\sigma_\mu\partial_\mu\lambda_L+\lambdab_R\sigmab_\mu\partial_\mu\lambda_R\right)
\end{eqnarray}
with $\sigma_\mu=(I,\sigma_i)$ and the ${\rm tr}$ symbol denotes the trace over the remaining $2\times 2$ blocks. 
The twisted fields transform under their internal symmetries according to
\begin{eqnarray}
\Psib_+ &\to& \Psib_+ G^\dagger\\\nonumber
\Psi_- &\to&  G\Psi_-\\\nonumber
\Lambda_- &\to& \Lambda_-H^\dagger\\\nonumber
\Lambda_+&\to& H\Lambda_+
\end{eqnarray}
In Weyl components the transformed matrix fields  read
\begin{eqnarray}
\Psib_+^\prime&=&\left(\begin{array}{cc} \psib_L G^\dagger & 0\\0 & \psib_R G^\dagger\end{array}\right)\quad \Lambdab_-^\prime=\left(\begin{array}{cc}0 & \lambdab_RH^\dagger\\ \lambdab_L H^\dagger& 0\end{array}\right)\nonumber\\
\Lambda_+^\prime &=&\left(\begin{array}{cc} H\lambda_R & 0 \\ 0 & H\lambda_L\end{array}\right)\qquad\Psi_-^\prime=\left(\begin{array}{cc}0 &G \psi_R\\ G\psi_L & 0\end{array}\right)
\label{chiral}
\end{eqnarray}

These internal symmetries can be made local by introducing appropriate covariant derivatives.
However, notice that unless $G=H$
it is not possible to write down gauge invariant mass terms which
are {\it simultaneously} twisted Lorentz invariant.\footnote{Gauge invariant bilinear mass terms which break
twisted Lorentz invariance but maintain 
the usual Lorentz symmetry are also possible. However, in section~\ref{global} and
section~\ref{gauge} we shall present 
evidence that condensates with this structure do not form in this lattice model.}. The natural
fermion bilinear invariant under the twisted Lorentz symmetry is
\beq {\rm Tr}\left(\Psib_+\Lambda_+ + \Lambdab_-\Psi_-\right)\eeq which clearly corresponds to
a set of degenerate Dirac mass terms when written out in components 
\beq
{\rm tr} \left(\psib_L\lambda_R+\psib_R\lambda_L+\lambdab_R\psi_L+\lambdab_L\psi_R\right)\eeq
Such a term breaks the original $G$ and $H$ symmetries  since it transforms as
\beq
\Psib_+G^\dagger H \Lambda_++\Lambdab_- H^\dagger G\Psi_-\label{axial}\eeq
Notice that if a condensate of this type forms in the theory it implies that the low energy theory
can be written in terms of Dirac spinors with the broken symmetries being realized
as axial transformations of these spinors.

From this analysis it should be clear that the fermionic interaction term that is invariant
under both gauge and twisted Lorentz symmetries
is  a four fermion operator of the form
\beq
{\rm Tr}\left(\Psib_+\Lambda_+\right){\rm Tr}\left(\Lambdab_-\Psi_-\right)\eeq
Notice that the explicit traces appearing in this operator act only on (twisted) spin and flavor
indices. Each fermion bilinear appearing in this term will also carry internal gauge indices which are suppressed
in this expression. Later we will assume that the gauge interactions factor into a strong
and weakly coupled sector and that the fermion bilinears appearing in the four fermi term
are singlets under the strong force but in general transform under the weak gauge
group. As can be seen from eqn.~\ref{axial} the four fermi term is nevertheless
fully gauge invariant. If a condensate of this form appears in the vacuum it
implies
that the bilinear terms themselves must be non-zero. 
Viewing the latter
as composite scalar fields it is clear that the formation of such a  four fermion condensate is equivalent to the statement that the
effective potential for the composite scalars must have developed
a minimum away from the origin and hence that the theory is in a Higgs phase. 
In addition, this four fermion operator simultaneously acts as an order parameter for the breaking of the {\it global}
symmetry of the twisted theory $SO(4)\times SO(4)\to SO(4)$ (each $SO(4)$ factor in the unbroken theory
corresponds to the usual
$SU(2)\times SU(2)$ chiral symmetry expected for two Dirac fermions). Notice that this global
symmetry of the twisted theory is much smaller than
the usual $SU(4)\times SU(4)$ chiral symmetry expected for four massless
Dirac fermions but follows once one gauges the kinetic terms for the projected fields differently\footnote{It is important to note that this $SO(4)$ flavor symmetry automatically enhances to
$SU(4)$ if $\Lambda_+$ and $\Psi_-$ carry identical representations of the internal symmetry group. In this case they can be thought of as arising by projection from a single
twisted field. This is the case for staggered quark simulations of QCD}
 
Finally, to complete the action of this continuum theory, we add the usual field strengths corresponding to
the gauge fields associated with the symmetries $G$ and $H$. 
In the next section  we will discretize this theory using the staggered fermion prescription and derive
a lattice model with similar properties. 

\section{Lattice model}
\label{lattice}
We start with the matrix fields $\Psi$ and $\Lambda$ introduced in the last section and
expand these matrices on a basis corresponding to
products of gamma matrices and associate these products with staggered fields. 
\begin{eqnarray}
\Psi(x)&=&\frac{1}{8}\sum_b \gamma^{x+b}\chi(x+b)\nonumber\\
\Lambda(x)&=&\frac{1}{8}\sum_{b} \gamma^{x+b}\xi(x+b) ,
\label{matrixfields}
\end{eqnarray}
where $\gamma^{x+b}=\prod_{i=1}^4 \gamma_i^{x_i+b_i}$ and the sums correspond to the
vertices in an elementary hypercube associated with lattice site $x$ 
as the components vary $b_i=0,1$.
It is easy to see that the projected matrix fields $\Psi_-$ and $\Lambda_+$ 
introduced in the continuum construction
corresponds to restricting the staggered field $\chi$ and $\xi$
to even and odd parity lattice sites. With a small abuse of notation we can write
\begin{eqnarray}
\Psib_+(x)&\to&\frac{1}{2}\left(1+\epsilon(x)\right)\chib(x)=\psib_+(x)\nonumber\\
\Lambdab_-(x)&\to&\frac{1}{2}\left(1-\epsilon(x)\right)\xib(x)=\lambdab_-(x)\nonumber\\
\Lambda_+(x)&\to&\frac{1}{2}\left(1+\epsilon(x)\right)\xi(x)=\lambda_+(x)\nonumber\\
\Psi_-(x)&\to&\frac{1}{2}\left(1-\epsilon(x)\right)\chi(x)=\psi_-(x)
\end{eqnarray}
where the parity of a lattice site is given by $\epsilon(x)=\left(-1\right)^{\sum_{\mu=1}^4 x_\mu}$.
The fields $\psi$ and $\lambda$ are termed  {\it reduced} staggered fermions since each contains half the
number of degrees of freedom of the usual staggered fermion and corresponds to
two rather than four Dirac fermions in the continuum limit \cite{redstag-Smit-1, Golterman:1984cy, redstag-Smit-2,  Catterall:2011ab}. Since the lattice
theory results from discretization of a continuum theory with equal numbers of left and right handed fields in any given representation of the gauge
group
it clearly does not violate the Nielsen-Niomiya theorem \cite{Nielson}.

The free continuum action given in eqn.~\ref{kinetic}  can then be 
recast as a lattice action by substituting the matrix expressions given in Eq.~(\ref{matrixfields}) into
the twisted action having replaced the derivative with a
symmetric difference operator and evaluating the 
trace using the orthogonality properties of the gamma matrices. The result is the
usual one expected for staggered fermions here written as the explicit sum of two {\it reduced}
staggered fermion actions
\begin{eqnarray}
&\sum_{x,\mu}&\eta_\mu(x)\psib_+(x)\left(\psi_-(x+\mu)-\psi_-(x-\mu)\right)\nonumber\\
&\sum_{x,\mu}&\eta_\mu(x)\lambdab_-(x)\left(\lambda_+(x+\mu)-\lambda_+(x-\mu)\right)
\end{eqnarray}
where the phase 
$\eta_\mu(x)$ is the usual staggered quark phase given by
\beq  \eta_{\mu}(x) = (-1)^{\sum_{i=1}^{\mu - 1} x_{i}} . \eeq
Just as before we can now take the staggered fields to transform
in different representations
of one or more symmetry groups. Following the continuum we take
\begin{eqnarray}
\psib_+ &\to& \psib_+ G^\dagger\\\nonumber
\psi_- &\to&  G\psi_-\\\nonumber
\lambdab_- &\to& \lambdab_-H^\dagger\\\nonumber
\lambda_+&\to& H\lambda_+
\end{eqnarray}
Again, these symmetries can be made local by inserting appropriate gauge links
between the $\psi$ and $\lambda$ fields on neighboring sites.
However, it is then impossible to write down a single site mass term
that preserves these symmetries.  If I take the matrix expression given in eqn.~\ref{matrixfields}
and substitute it into the expression for the mass term one
easily derives the usual single site staggered mass term
\beq
\left(\psib_+(x)\lambda_+(x)+\lambdab_-(x)\psi_-(x)\right)\eeq
which, in general, is no longer gauge invariant.

However, it is possible to write down a gauge invariant lattice four fermion term by analogy
with the continuum. As in our discussion of the twisted continuum theory  we will assume
that the interactions factorize into a strongly coupled sector and a weakly coupled sector and that
the $\psi$ and $\lambda$ fields transform differently under the weak symmetries.
We can then introduce composite scalar fields which are singlets under the strong gauge symmetry
but transform under the weak group:
\begin{eqnarray}
u_+(x)&=&\psib_+(x)\lambda_+(x)\nonumber\\
u^\dagger_-(x)&=&\lambdab_-(x)\psi_-(x)
\end{eqnarray}
A gauge invariant four fermi term can then be constructed by connecting these composite scalar
fields defined in neighboring lattice sites
through an appropriate set of weak gauge links.

As in the continuum the formation of a condensate of this form signals a Higgsing of
lattice gauge symmetry. But as we saw earlier this symmetry has an axial
character in the usual (untwisted) continuum theory since it acts differently on left and right
handed components of a given Dirac spinor.
This is at first sight puzzling; how can a manifestly vector
lattice symmetry be reconciled with an axial symmetry in the continuum limit ?
The resolution to this
puzzle is easily found; recall that the
even parity lattice fields $\lambda_+$ are contained  in the continuum twisted matrix field $\Lambda_+$ 
and transform according to $H$ while
odd parity lattice fields $\psi_-$ live in $\Psi_-$ and transform according to $G$.
The condensate we examine couples $\Psib_+$ to $\Lambda_+$ and contains
terms like $\psib_L\lambda_R+\lambdab_R\psi_L$. This in turn implies that the
physical fields surviving at long distance comprise Dirac spinors whose chiral
components transform differently under these broken internal symmetries.
This chiral or axial  behavior cannot however be transferred to
the lattice since the
natural candidate for a lattice Dirac spinor would be assembled from $\psi$ and $\lambda$ fields
at different sites. Such an object
does not transform 
covariantly under lattice gauge transformations and hence no exact axial symmetry
is possible in the lattice model. However, formally these constraints disappear at vanishing
lattice spacing.

Notice that
the single site condensate discussed here is
similar to that encountered in conventional staggered lattice simulations of QCD. 
The usual
$U(1)_V\times U(1)_A$ of staggered fermions is then equivalent to a global $U(1)\times U(1)$ vector symmetry of 
the system of two reduced staggered fermions that we study. The condensate breaks this
symmetry down to its diagonal $U(1)$ subgroup which then
becomes fermion number. The orthogonal combination is then to be
interpreted as the usual broken axial symmetry of staggered fermions. The reader should
notice the structure of this argument; a vector symmetry of the system of 2 reduced staggered
fermions is reinterpreted as a (broken) axial symmetry in the usual staggered fermion
model.

Finally, as for the continuum, we will need to introduce kinetic terms for the lattice
gauge fields by 
adding the usual Wilson plaquette terms built from the corresponding gauge links.

We will illustrate these ideas by constructing technicolor-like models
in which the gauge interactions
factorize into strong and weakly coupled sectors with corresponding
gauge groups $S=SU(N)$ and $W=SU(M)$ respectively. 
As an explicit example consider the case where the twisted matrix fermion $\Psi_-$ transforms in the fundamental representation
of both $S$ and $W$ while $\Lambda_+$  transforms as a fundamental under $S$ but is sterile under $W$.
Thus any bilinear of the form $\Tr\left(\Psib_+\Lambda_+\right)$ transforms as a anti-fundamental
under $W$ and hence a condensate with this structure will break $SU(M)\to SU(M-1)$.

\section{Global symmetry breaking}
\label{global}

Before turning to the problem of breaking local or gauge symmetries it is instructive to first
consider the
simpler case where the would be broken symmetry is global. It is well
know how to search for spontaneous symmetry breaking in this case. One measures the expectation value of
some suitable order parameter in a modified version of the theory which incorporates a symmetry
breaking external field coupled to that order parameter.
The symmetry breaking is then signaled by a nonzero value for the expectation value of
the order parameter
as the external field is sent to zero {\it after} the thermodynamic limit is taken.
In this case the natural order parameter is the condensate $<\psib\lambda>$.

We can look for such a condensate using numerical simulations of
the associated  model based on reduced staggered fermions  as
described in section~\ref{lattice}. The gauging of
the lattice kinetic term is then given explicitly as
\begin{widetext}
\begin{eqnarray}
&\sum_{x,\mu}&\psib_+(x)\left(U_\mu(x)V_\mu(x)\psi_-(x+\mu)-U^\dagger_\mu(x-\mu)V^\dagger_\mu(x-\mu)\psi_-(x-\mu)\right)\nonumber \\
&\sum_{x,\mu}&\lambdab_-(x)\left(V_\mu(x)\lambda_+(x+\mu)-V^\dagger_\mu(x-\mu)\lambda_+(x-\mu)\right)
\label{staggered_kinetic}
\end{eqnarray}
\end{widetext}
where we set the weak gauge links $U_\mu(x)=1$ for the case of a global symmetry. 
As an external symmetry breaking perturbation we add to the action the following term
\beq
\sum_x g\left[\phi\psib_+(x)\lambda_+(x)+\phi^*\psi_-(x)\lambdab_-(x)\right]\eeq
with $\phi$ a constant external field transforming in the fundamental of the $SU(M)$ group.  We break
the $SU(2)$ symmetry explicitly by setting $\phi=(1,0)$ and in this case the Yukawa coupling $g$ gives the
magnitude of the symmetry breaking external field.

We have studied this model for the case $N=3$ and $M=2$ using
the standard RHMC algorithm and working in the phase quenched approximation (the reduced staggered fermion
models we consider here  in general have a sign problem)
\begin{figure}
\begin{center}
\includegraphics[height=62mm]{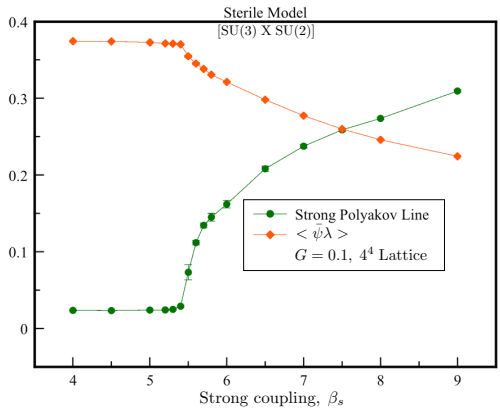}
\caption{\label{compare-global} Condensate $<\psi\lambda>$ and $SU(3)$ Polyakov line $|P|$
vs $\beta_S$ for fixed $g=0.1$ and $L=4$}
\end{center}
\end{figure}
Fig.~\ref{compare-global} shows the results for the bilinear condensate $<\psib\lambda>$ as the strong gauge coupling $\beta_S$ is
varied
for a  fixed value of the external field $g=0.1$
on a lattice of size $L=4$. On the same plot we show the Polyakov line
corresponding to the strongly coupled $SU(3)$ gauge field. The latter is small at strong coupling (small $\beta_S$) and rises
for $\beta_S>5.5$ corresponding to the onset of a deconfinement in the $SU(3)$ sector for weak enough coupling. The condensate shows
a complementary behavior saturating at a non-zero value for small $\beta_S$ and dropping quickly for large
values of $\beta_S$. Indeed, one can see that the change in behavior for both these quantities is
highly correlated; as soon as the strong sector of the model deconfines the condensate starts to fall. The rate at which the
condensate falls to zero is governed by the magnitude of the symmetry breaking field - the plot shows data for $g=0.1$ but
we have observed that the fall off with $\beta_S$ appears more abrupt for smaller $g$. Notice that the effective number of fermion flavors in these
simulations  is given by $N_f=6=2\times (2_{\rm SU(2)}+1_{\rm sterile})$ since each reduced field yields two
continuum  Dirac fermions.
We expect confinement for this number of flavors in $SU(3)$.
\begin{figure}
\begin{center}
\includegraphics[height=62mm]{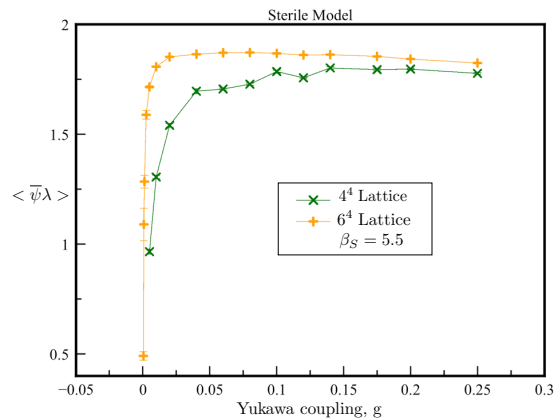}
\caption{\label{globalcondvsg} Condensate $<\psi\lambda>$ vs $g$ at $\beta_S=5.5$ on lattices of size $L=4$ and $L=6$}
\end{center}
\end{figure}

Of course the question of whether we are seeing spontaneous symmetry breaking hinges on the behavior of
the condensate  as we send the external field $g$ to
zero.
Fig.~\ref{globalcondvsg} shows the condensate for lattices of size
$L=4$ and $L=6$ as we scan in the symmetry breaking field $g$ at a fixed value of
$\beta_S=5.5$ representative of the scaling window for these small lattices. A broad plateau is visible which declines slowly
for large $g$ and turns over to approach zero for very small $g$. This turnover point moves to
smaller $g$ as the lattice size is increased. This result is consistent with the expected behavior for a system in which symmetry
is spontaneously broken; while the order parameter must vanish on any finite
system one can obtain a nonzero value at ever decreasing values of
the external field as the volume of the system increases. In principle one should perform an extrapolation of the condensate to
infinite volume and then examine how this extrapolated value varies as the external field is sent to zero. This is beyond the scope
of this initial work but would be required in any followup study. However, the behavior we see is certainly consistent with
existence of a non-zero condensate in the thermodynamic limit.

\begin{figure}
\begin{center}
\includegraphics[height=62mm]{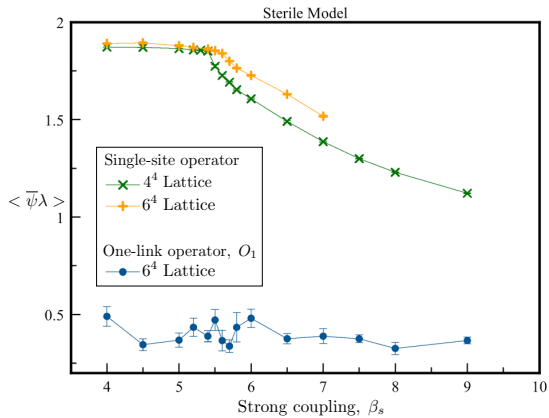}
\caption{\label{globalcondvsbeta} Condensate $<\psi\lambda>$ vs $\beta_S$ for fixed $g=0.1$ ($L=4$) and $g=0.02$ ($L=6$)}
\end{center}
\end{figure}

We have also checked the volume dependence of the condensate; fig.~\ref{globalcondvsbeta} shows data for both
$L=4$ and $L=6$.  Notice that we have scaled the external field
as the inverse of the lattice volume to produce these curves. This is the scaling expected for systems with
chiral symmetry breaking where the lattice condensate is expected to be a function of $gV$.
The larger lattice data then lies close to the results for $L=4$. 
Deviations from this naive scaling likely reflect both finite volume and lattice spacing (running) effects.
Notice that the data collapse closely on a single curve in the confined phase while the
deviation is larger in the deconfined phase. We think that the former likely indicates the 
effects of the running coupling while the latter effect is likely due to finite volume. To test these
ideas would require simulations on larger volumes and for smaller lattice spacings than in
our current study and we hope to pursue this in future work.

The circles visible in this
figure correspond to the measurement of an exactly gauge invariant condensate based on the one link
operator 
\beq O_1=\psib_+(x)V_\mu(x)\psi_-(x+\mu)\eeq
 In terms of the twisted theory this corresponds to an operator of the
form $\Tr\left(\Psib_+\Psi_-\gamma_4\right)$. While this term is invariant under the usual Lorentz symmetry it is
clearly {\it not} invariant under the twisted Lorentz symmetry. The numerical results that are shown correspond to the Monte Carlo
average of the absolute value of this operator measured on a given configuration (the naive average is statistically consistent with zero).
The numerical results show that the one-link condensate is much smaller than the single site condensate and, perhaps more importantly, shows little
dependence on the magnitude of the strong coupling $\beta_S$.  

Actually, this should not be surprising. Once the weak interactions are set to zero   one can
interpret the field content as corresponding to a single conventional staggered
fermion built from $(\psi,\lambdab^1)$ with conventional single site mass term together
with an additional massless reduced staggered fermion $\lambda^2$ where
the explicit indices correspond to the exact weak $SU(2)$ symmetry. The condensate we
observe is then just the usual single site condensate
expected for staggered fermions. Following our experience with QCD we expect that this survives to the continuum limit
and breaks all the axial symmetries of the theory which must hence include this exact $SU(2)$.

Another way to understand this result is to note that the single site condensate is
the natural object to arise in the twisted construction since it is twisted Lorentz
invariant.
The situation is also similar to recent work on lattice ${\cal N}=4$ super Yang-Mills
\cite{Catterall:2009it,susypapers}. In this
case it is clear that the continuum limit of the supersymmetric lattice model is the twisted Yang-Mills theory and one has to
work harder to show that one can {\it untwist} the model to target the usual ${\cal N}=4$ theory with $SU(4)$
R symmetry. Indeed, in this case, a proof that the global symmetries of the theory enhance to the full
$SU(4)$ symmetry is equivalent to a proof that all the supersymmetries in the theory are restored without
fine tuning. A necessary condition for this to occur is that the theory remain massless and no condensates form. While
this is quite possibly the case in ${\cal N}=4$ Yang-Mills it does not appear to be true in the theory considered here.

To summarize: the numerical results in this section
lend strong support to the idea that this staggered fermion system spontaneously breaks
an exact global $SU(2)$ symmetry as a result of strong gauge dynamics. 
Under rather general assumptions such a system
should therefore exhibit a Higgs mechanism once this global symmetry is gauged. We will show direct evidence for this
in the next section.

\section{Gauge symmetry breaking}
\label{gauge}

The breaking of symmetries  in a gauge theory is somewhat
more subtle; in this case the {\it global} external field must be replaced by a {\it local} field $\phi(x)$ which also transforms under
the gauge symmetry in such a way that
this term is exactly gauge invariant.  This requires that $\phi$ transforms as a fundamental of $SU(2)$
\beq
\phi(x)\to W(x)\phi(x)\eeq
where $W$ corresponds to weak gauge transformations.
To render the path integral well defined after
integration over $\phi(x)$ one must
then also add a suitable action for $\phi(x)$. We choose an additional simple term  $\sum_x \phi^\dagger(x)\phi(x)$.
The effect of these Yukawa  terms is to add a small  gauge invariant  four fermion interaction to the action that
favors the conjectured symmetry breaking pattern. 
In addition we now allow the lattice gauge field $U_\mu(x)$ appearing in eqn.~\ref{staggered_kinetic} to fluctuate
and add an associated Wilson plaquette term
to the action.

As in the case
of a global symmetry it will be important in our
argument to show that
the magnitiude of any induced four fermion condensate is insensitive to the value of this coupling $g$ and depends
instead only on the magnitude of the strong gauge coupling. Indeed,
following the same strategy used in the case of a global symmetry, 
we will consider the limit $g\to 0$
{\it after} the thermodynamic limit.

To distinguish such a Higgsed phase from the weakly coupled one 
one would like to have an order parameter.
Elitzur's theorem tells us that the vacuum expectation value of
any gauge {\it variant} local observable will vanish even for non-zero $g$ if 
the action is gauge invariant \cite{Elitzur:1975im}. 
Instead one should distinguish a Higgs phase by analysing the
asymptotic behavior of non local quantities such as Wilson loops or Polyakov lines. 
In our numerical work we will base our conclusions 
partly on the behavior of the Polyakov line constructed from the
weak gauge field. Since this quantity represents the contribution to the
free energy of a fermion charged under the gauge symmetry we expect to see
a change in this quantity  as the system enters a Higgs phase. We have also
monitored a gauge invariant four fermion condensate which in the continuum limit 
represents the absolute square of the bilinear condensate studied in the case of a global symmetry.
\beq
\sum_\mu
<\psib_+(x)\lambda_+(x)>U_\mu(x)<\lambdab_-(x+\mu)\psi_-(x+\mu)>\eeq
While this quantity is not strictly an order parameter we might expect
it to be strongly enhanced in a Higgs phase. It is important to notice that
the fermion bilinears appearing in the above Wick contractions are singlets under the strong
gauge interactions and carry only weak gauge indices. 
A final gauge invariant expression is then constructed by tying these
objects together using the weak gauge link $U_\mu$.
\begin{figure}
\begin{center}\includegraphics[height=62mm]{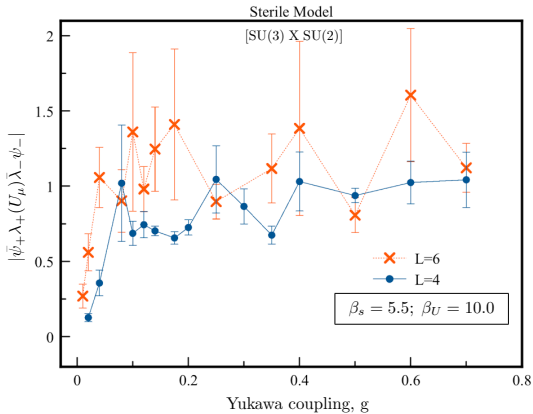}\end{center}
\caption{\label{fig1}Absolute value of the four fermion condensate vs Yukawa coupling $g$}
\end{figure}
Again we have tested these ideas for
the model with $N=3$ and $M=2$ where one
expects complete breaking of the weak $SU(2)$ for sufficiently strong $\beta_S$.
Figure~\ref{fig1} shows
a plot of the four fermion condensate versus the auxiliary Yukawa coupling for $\beta_S=5.5$ and $\beta_W=10.0$.
Similar to the case where the $SU(2)$ symmetry is global we see that
the magnitude of the condensate remains approximately constant as we
decrease $g$ until very small $g$ and that this effect is enhanced as the volume increases.

\begin{figure}
\begin{center}\includegraphics[height=62mm]{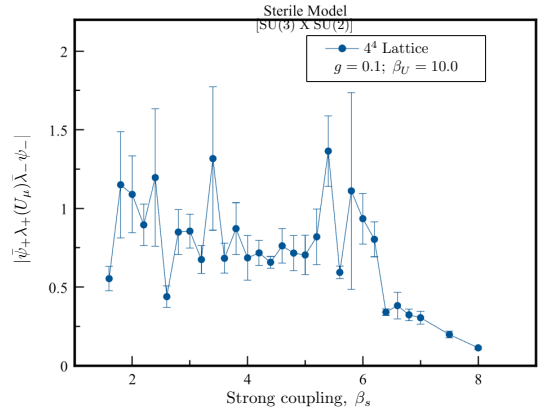}\end{center}
\caption{\label{fig4} Absolute value of the four fermion condensate vs $\beta_S$}
\end{figure}
To verify that the appearance of this condensate is a direct result of the strong $SU(3)$ dynamics
we have also examined the four fermion condensate as a function of $\beta_S$ for fixed auxiliary
coupling $g=0.1$ and $L=4$. The results are
shown in figure~\ref{fig4}.
\begin{figure}
\begin{center}\includegraphics[height=62mm]{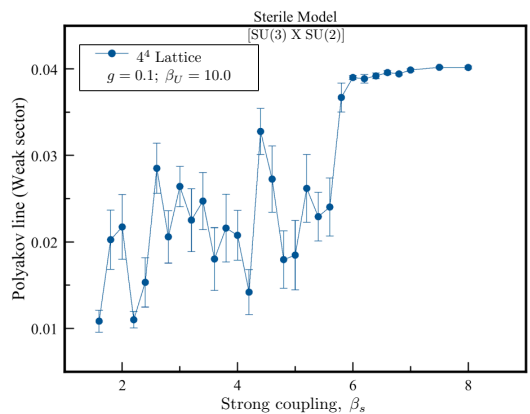}\end{center}
\caption{\label{poly}Weak Polyakov line versus $\beta_S$}
\end{figure}
Clearly the condensate is enhanced at strong coupling falling to small values as $\beta_S\to\infty$. Most
importantly we see no sign of a phase transition as we vary the strong coupling. This is evidence that
the Higgs phase of the lattice theory may survive the continuum limit $\beta_S\to \infty$.

We can see a
signal of the appearance of a Higgs phase even more clearly by looking at the plot in fig.~\ref{poly} of
the Polyakov line corresponding to the {\it weak} gauge field as we
vary the {\it strong} gauge coupling $\beta_S$. For large $\beta_S$ the weak Polyakov line $P_W$ is large
but as $\beta_S$ is lowered it rapidly crosses over to fluctutate around a much 
smaller but non zero value. Since the weak Polyakov line $P_W=e^{-FT}$ measures the free energy $F$ of
an isolated static quark in the fundamental representation of the weak gauge group a value approaching unity is
associated with a deconfined phase. This is to be expected for a bare weak gauge coupling 
$\beta_W=10.0$ on such a small lattice. Conversely, a confining phase typically
would be associated with a small value of $P_W$ (see the $SU(3)$ Polyakov line data in fig.~\ref{compare-global}). 
What is observed for small $\beta_S$ is intermediate between these two regimes and may be a signal
of a Higgs phase.
Furthermore, the crossover between these two regimes corresponds precisely
to the switching on of the four fermion condensate and the observation of
deconfinement in the strong sector.
Thus our numerical
results are at least consistent with the dynamical
generation of a non-zero four fermion condensate and the appearance of
a Higgs phase as a result of
strongly coupled dynamics. Clearly to be sure of this interpretation requires much larger simulation volumes and
an analysis of the full static quark potential which we hope to turn to later.

\begin{figure}
\begin{center}\includegraphics[height=62mm]{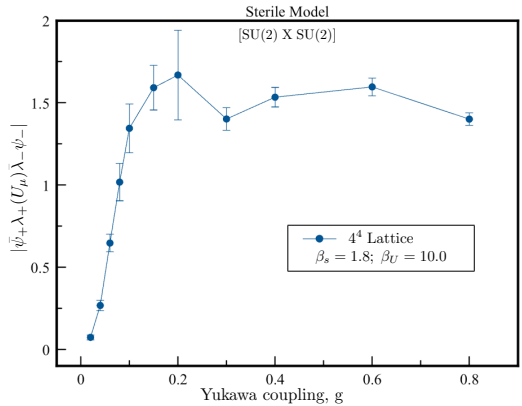}\end{center}
\caption{\label{su2} Absolute value of the four fermion condensate  vs $g$ for an $SU(2)$ strong sector}
\end{figure}

At this point it is legitimate to ask whether the effect we are seeing is any way connected to the
use of the phase quenched approximation. We can answer this by considering results from
the same model when we employ an $SU(2)$ strong sector. Figure.~\ref{su2} shows a plot of
the condensate versus auxiliary coupling at $\beta_S=1.8$ (see \cite{Catterall:2011ab} for
details on this choice of coupling) and $\beta_W=10.0$ for this
model. Since $SU(2)$ is pseudoreal the fermion measure for this theory is positive definite for
vanishing auxiliary coupling. Clearly the condensate still rises to a non-zero
plateau  even at small $g$ where
we have observed that the phase is negligible.

\section{Breaking to a diagonal subgroup}

It is possible to explore other models in an effort to understand
how generic is this symmetry breaking scenario. One simple generalization of
the sterile model discussed earlier  
again assumes that $\psi$ and $\lambda$ transform in the fundamental 
representation of a strong
$SU(N)$ interaction. But now the weak sector has an $SU(M)\times SU(M)^\prime$
structure with the fermions $\psi$ and $\lambda$ transforming in the
$({\bf 1}, {\bf \Box})$ and $({\bf \Box},{\bf 1})$ representations of this product group.
The kinetic terms are now given explicitly as
\begin{widetext}
\begin{eqnarray}
&\sum_{x,\mu}&\psib_+(x)\left(U_\mu(x)V_\mu(x)\psi_-(x+\mu)-U^\dagger_\mu(x-\mu)V^\dagger_\mu(x-\mu)\psi_-(x-\mu)\right)\nonumber \\
&\sum_{x,\mu}&\lambdab_-(x)\left(W_\mu(x)V_\mu(x)\lambda_+(x+\mu)-W^\dagger_\mu(x-\mu)V^\dagger_\mu(x-\mu)\lambda_+(x-\mu)\right)
\end{eqnarray}
\end{widetext}
where the gauge field
$V_\mu(x)$ corresponds to the strongly coupled $SU(N)$ sector while $U_\mu(x)$ and $W_\mu(x)$ are the
gauge fields for the two independent $SU(M)$ symmetries. It is clear that a non-zero condensate will now indicate
a breaking of the two $SU(M)$'s down to their diagonal $SU_D(M)={\rm diag}\left(SU(M)\times SU(M)^\prime\right)$.
We will focus on the case $N=3$ and $M=2$ so that the weak symmetries are just $SU(2)\times SU(2)$.

Let us first examine the case where the weak gauge coupling is set to zero and the weak symmetry is purely global.
Since each reduced staggered field contributes 2 Dirac fermions in the continuum
limit  the theory will contain eight Dirac fields
with global symmetry $SU_V(8)\times SU_A(8)$.  As a result of strong interactions
we expect that this symmetry will break according to the pattern
\beq
SU_V(8)\times SU_A(8)\to SU_V(8). 
\label{su8}
\eeq
The $SU(2)\times SU(2)$ weak symmetries of the staggered lattice theory 
constitute a subgroup of this continuum global symmetry. The precise embedding of these
weak symmetries in the continuum limit can be obtained by the following argument
\footnote{We thank Maarten Golterman and Yigal Shamir for
pointing out this argument and sharing their notes with us \cite{MS}}.

We start by assuming that each staggered field yields 2 Dirac fermions in the
continuum eg. $\psi_1\to \{\Psi_1^1,\Psi_1^2\}$ with the upper indices reflecting this extra factor of two. We can then arrange these
continuum fields into an eight component vector
\beq\left(\begin{array}{cccccccc}
\Psi^1_1 & \Lambda^1_1 & \Psi^1_2 & \Lambda^1_2 &
\Psi^2_1 & \Lambda^2_1 & \Psi^2_2 & \Lambda^2_2 \end{array}\right)\eeq
In this representation the {\it broken}  generators $\tau_a =\tau^1_a-\tau^2_a,\,a=1\ldots 3$ take the explicit form
\beq
\begin{array}{ccc}
\tau_1&=\left(\begin{array}{cc}
0&\sigma_3\\
\sigma_3&0\end{array}\right)\times I_2\qquad &
\tau_2=\left(\begin{array}{cc}
0&i\sigma_3\\
-i\sigma_3&0\end{array}\right)\times I_2 \\
\tau_3&=\left(\begin{array}{cc}
\sigma_3&0\\
0&-\sigma_3\end{array}\right)\times I_2
\end{array}
\eeq
where the two dimensional unit matrix $I_2$ represents this extra factor
of two degeneracy associated to the upper indices of $\Psi,\Lambda$. 
We will neglect the $I_2$ factor in what follows since it enters trivially in our analysis.
On the lattice the usual single site fermion condensate takes the form
\beq
\sum_{a=1}^2\psib^a_+\lambda^a_++\lambdab^a_-\psi^a_-
\label{cond}\eeq
corresponding to two independent staggered fermion condensates. This clearly
breaks the lattice $SU(2)\times SU(2)$ symmetry down to its diagonal subgroup. 
We then expect that this breaking pattern corresponds to a continuum condensate of the form
\beq
\Sigma=\left(\begin{array}{cc}
\sigma_1&0\\
0&\sigma_1\end{array}\right)\times I_2
\eeq
Standard universailty arguments tell us that it should be possible to change basis for our fermion fields to force this condensate to take
the canonical flavor symmetric form $\Sigma=I_8$. To accomplish this first diagonalize $M\to P^\dagger MP$ using the
unitary transformation
\beq
P=\left(\begin{array}{cc}
\frac{1}{\sqrt{2}}\left(\sigma_3+\sigma_1\right)& 0\\
0& \frac{1}{\sqrt{2}}\left(\sigma_3+\sigma_1\right)\end{array}\right)\eeq
This results in a condensate of the form
\beq
\Sigma^\prime=\left(\begin{array}{cc}\sigma_3&0\\
0&\sigma_3\end{array}\right)\eeq
To transform this to the unit matrix we employ the non-anomalous chiral transformation $M^\prime \to QM^\prime Q$
with
\beq
Q=\left(\begin{array}{cc}
D&0\\
0&D^\dagger\end{array}\right)\eeq
where the $2\times 2$ matrix
 \beq
D=\left(\begin{array}{cc}
1&0\\
0& i\gamma_5\end{array}\right)\eeq
To find the explicit form of the broken generators in this new basis we transform them according to
the rule $Q^\dagger P^\dagger \tau_a PQ$.  It is straightforward to show that
the broken generators acquire an {\it axial} character in the new basis (we have reinserted the $I_2$ at this point)
\begin{widetext}
\beq
\begin{array}{ccc}
\tau^\prime_1=\gamma_5\left(\begin{array}{cc}
0&-i\sigma_1\\
i\sigma_1&0\end{array}\right)\times I_2\qquad &
\tau^\prime_2=\gamma_5\left(\begin{array}{cc}
0&\sigma_1\\
\sigma_1&0\end{array}\right)\times I_2\qquad &
\tau^\prime_3=\gamma_5\left(\begin{array}{cc}
\sigma_2&0\\
0&-\sigma_2\end{array}\right)\times I_2
\end{array}
\eeq
\end{widetext}
This confirms our earlier arguments that the broken generators do indeed correspond to axial symmetries in the continuum limit.
We see that the staggered fermion action we use picks out a particular breaking direction corresponding to a specific embedding of the weak symmetries
into the global symmetry group. The fact that the broken symmetries are axial in the continuum limit is
a necessary condition, according to
the Vafa-Witten theorem, for this broken phase of
the lattice theory to survive the continuum limit.

As in the sterile case we have checked these ideas using explicit simulations.  In this case we utilize an auxiliary scalar field which is a bi-fundamental 
in the two weak groups. For a scan in the auxiliary coupling $g$ we again
set the
strong gauge coupling $\beta_S=5.5$ corresponding to a
confining regime of the $SU(3)$ gauge interactions
and use weak couplings $\beta_U=\beta_W=10.0$. As before these results were obtained in the phase quenched
approximation. 
\begin{figure}
\begin{center}
\includegraphics[height=62mm]{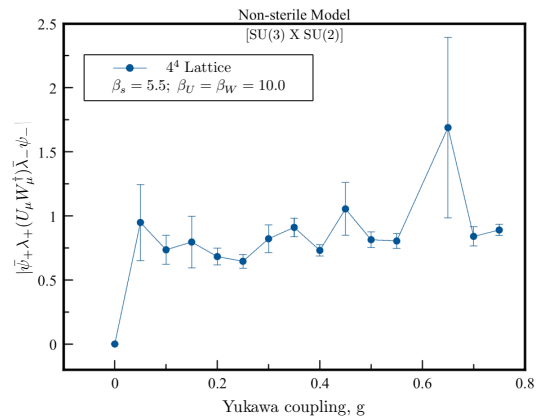}\end{center}
\caption{\label{nonsterile1} Absolute value of the four fermion condensate vs $g$ with $\beta_{S} = 5.5$ for the case
of breaking to the diagonal subgroup}
\end{figure} 
\begin{figure}
\begin{center}
\includegraphics[height=62mm]{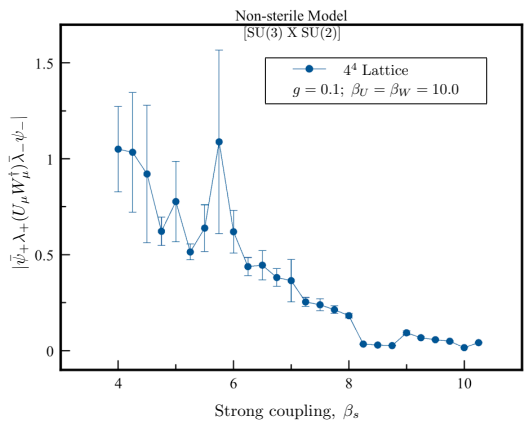}\end{center}
\caption{\label{nonsterile2} Absolute value of the four fermion condensate vs $\beta_{S}$ for auxiliary Yukawa coupling g = 0.1
for the case of breaking to a diagonal subgroup}
\end{figure} 
The plots (figs.~\ref{nonsterile1} and \ref{nonsterile2}) show, once again, that a nonzero four fermion condensate develops,
which is insensitive to the auxiliary Yukawa coupling $g$, and whose magnitude is determined 
by the strong gauge coupling $\beta_S$. We observe that it falls towards zero once the $SU(3)$ sector
deconfines.\footnote{One might wonder whether one could break a $U(1)\times U(1)\to U(1)$ gauge symmetry along these lines.
However, the measure for a single reduced staggered field is not invariant under a local
$U(1)$ transformation and so this lattice symmetry suffers from an anomaly.}

\section{Discussion}

We have shown that lattice theories comprising two {\it reduced} staggered fermion fields 
can be constructed in such a way that they can naturally
generate non-perturbative condensates that break
exact weakly coupled gauge symmetries
as a result of strongly coupled gauge dynamics. The key element that allows for a Higgs
mechanism to operate 
in these models is the absence of
single site fermion bilinears which are invariant under the gauge symmetries. This implies that
the first gauge invariant object that can condense is 
a four fermion operator. Remarkably,  these broken
symmetries, which start out as vector symmetries in the
staggered lattice theory, can be identified with axial symmetries in the continuum limit and the result is hence compatible with the
Vafa-Witten theorem.
The vector-like nature of the continuum theory ensures that it is 
 trivially free of gauge anomalies \cite{KS}.

These features arise as a consequence of the fact that
the lattice theory is obtained by discretization
of a continuum theory in which the original Lorentz symmetry has been twisted
with an internal chiral-flavor symmetry. The free action of these twisted theories
then naturally decomposes into two independent components which may be gauged
independently. The resultant models, while having a vector like
field content, can act nevertheless like chiral gauge theories by not admitting
mass terms that are simultaneously both gauge invariant and {\it twisted} Lorentz invariant. 

While preparing this paper we discovered a similar construction had been proposed  many years
ago by Banks et al. \cite{Banks:1982gt} for the case of breaking to a
diagonal subgroup. This earlier work employed arguments based on a strong coupling
expansion to argue for a Higgs phase at non-zero lattice spacing. Our numerical work provides evidence that
this Higgs phase may survive into the continuum limit which is also consistent with our interpretation that
the broken
symmetries behave as axial symmetries in that limit.

In the continuum limit the formation of such a  condensate will break
the usual global axial symmetries in addition to any internal symmetries.
Thus the  continuum theory naturally contains additional light
Goldstone bosons in addition to any massive gauge bosons
associated with the breaking of gauge symmetry. 
In addition, one expects that any
fermions participating in the condensate gain mass on the order of the symmetry breaking
scale and decouple from the low energy spectrum.

In conclusion,
it seems that the lattice models discussed here may serve as useful toy models
for understanding the possibilities for dynamical symmetry breaking in strongly
coupled gauge theories and can be used
to test ideas such as the maximal attractive channel hypothesis 
and tumbling scenarios. In many respects
the behavior of these theories mimicks that of chiral fermions although in the
continuum limit they more closely resemble left-right symmetric models
such as Patti-Salam \cite{Patti}. It would be interesting
to investigate such models via strong coupling expansions which would avoid possible
sign problems.

\section*{Acknowledgments}
The authors would like to thank Jay Hubisz and Anna Hasenfratz for comments on
an early draft of the manuscript and Poul Damgaard, in particular, for alerting us to the
earlier paper \cite{Napoly} and for extensive and extremely useful discussions on the material
presented in this paper. SMC would also like to thank the Aspen Institute for Physics
for hospitality during the completion of this work and the DOE for partial support
under grant DE-FG02-85ER40237. The simulations were carried out using USQCD resources
at Fermilab.


\begin{thebibliography}{99}

\bibitem{TC-intro}
S.~Weinberg, ``Implications of dynamical symmetry breaking,'' {\em Phys.Rev.D},
  vol.~13, p.~974, 1976.

\bibitem{TC-intro2}
L.~Suskind, ``Dynamics of spontaneous symmetry breaking in the weinberg-salam
  theory,'' {\em Phys.Rev.D}, vol.~20, p.~2619, 1979.

\bibitem{Raby:1979my} 
  S.~Raby, S.~Dimopoulos and L.~Susskind,
  ``Tumbling Gauge Theories,''
  Nucl.\ Phys.\ B {\bf 169}, 373 (1980).
 
\bibitem{gut}
H. Georgi and S. L. Glashow,
``Unity of Elementary Particle Forces", Phys. \ Rev. \ Lett {\bf 32}, 438 (1974)

\bibitem{Elitzur:1975im} 
  S.~Elitzur,
  ``Impossibility of Spontaneously Breaking Local Symmetries,''
  Phys.\ Rev.\ D {\bf 12}, 3978 (1975).

\bibitem{Vafa:1984xg} 
 C.~Vafa and E.~Witten,
``Parity Conservation in QCD,''
Phys.\ Rev.\ Lett.\  {\bf 53}, 535 (1984).

\bibitem{Poppitz}  ``Chiral Lattice Gauge Theories via Mirror Decoupling: A 
Mission (Im)possible ?", E. Poppitz and Y. Shang, Int. J. Mod. Phys A25:2761, 2010.

\bibitem{Luescher}  ``Chiral gauge theories revisited", M. L\"{u}scher, lectures at Erice (2000)
hep-th/0102028.

\bibitem{Golterman}  ``Lattice chiral gauge theories", Maarten Golterman, Nucl. Phys. Proc. Suppl. 94 (2001) 189.
hep-lat/0011027.

\bibitem{Kikukawa} ``Domain wall fermion and chiral gauge theories on the lattice with exact gauge invariance", Y. Kikukawa, Phys. Rev. D65 (2002) 074504.



\bibitem{Catterall:2009it} 
  S.~Catterall, D.~B.~Kaplan and M.~Unsal,
  ``Exact lattice supersymmetry,''
  Phys.\ Rept.\  {\bf 484}, 71 (2009)
  [arXiv:0903.4881 [hep-lat]].


\bibitem{redstag-Smit-1}
C.~V. den Doel and J.~Smit, ``Dynamical symmetry breaking in two flavor su(n)
  and so(n) lattice gauge theories,'' {\em Nuclear Physics B}, vol.~228, no.~1,
  pp.~122 -- 144, 1983.
  

\bibitem{Golterman:1984cy} 
  M.~F.~L.~Golterman and J.~Smit,
  ``Selfenergy and Flavor Interpretation of Staggered Fermions,''
  Nucl.\ Phys.\ B {\bf 245}, 61 (1984).

\bibitem{redstag-Smit-2}
W.~Bock, J.~Smit, and J.~C. Vink, ``Fermion-higgs model with reduced staggered
  fermions,'' {\em Phys.Lett.B}, vol.~291, p.~297, 1992.
  
\bibitem{Catterall:2011ab} 
  S.~Catterall, R.~Galvez, J.~Hubisz, D.~Mehta, A.~Veernala and ,
  ``Non-abelian gauged NJL models on the lattice,''
  Phys.\ Rev.\ D {\bf 86}, 034502 (2012)
  [arXiv:1112.1855 [hep-lat]].

\bibitem{Nielson}
H. Nielson and Ninomiya, `` No go theorem for regularizing chiral fermions", Phys. Lett. B105, 219 (1981)

\bibitem{Napoly}
O. Napoly, ``A Lattice Model for the dynamical Higgs mechanism", Phys. Lett. B159 (1985) 353.

\bibitem{susypapers}
S. Catterall, J. Giedt and A. Joseph, ``Twisted supersymmetries in lattice Yang-Mills theory",
 arXiv:1306.5668.

\bibitem{MS} Maarten Golterman and Yigal Shamir private communication.
\bibitem{KS}
  L.~H.~Karsten and J.~Smit,
  ``Lattice Fermions: Species Doubling, Chiral Invariance, and the Triangle Anomaly,''
  Nucl.\ Phys.\ B {\bf 183}, 103 (1981).

\bibitem{Banks:1982gt} 
  T.~Banks and A.~Zaks,
  ``Chiral Analog Gauge Theories on the Lattice,''
  Nucl.\ Phys.\ B {\bf 206}, 23 (1982).

\bibitem{Patti} J. Patti and A. Salam, 
``Lepton  number as the fourth color",
Phys. \ Rev. {\bf D10} (1974) 275.



 \end{thebibliography}
\end{document}